\documentclass{emulateapj}
\journalinfo{Accepted for publication in the Astrophysical Journal Letters}

\newcommand{\av}{\ensuremath{A_{V}}}
\newcommand{\nh}{\ensuremath{N_{\rm{H}}}}
\newcommand{\ktFT}{\ensuremath{k_{\rm{B}}T}}
\newcommand{\pwFP}{\ensuremath{\Gamma}}
\newcommand{\fxFT}{\ensuremath{F_{\rm{X}}^{\rm{(th)}}}}
\newcommand{\fxFP}{\ensuremath{F_{\rm{X}}^{\rm{(pl)}}}}
\newcommand{\lxFT}{\ensuremath{L_{\rm{X}}^{\rm{(th)}}}}
\newcommand{\lxFP}{\ensuremath{L_{\rm{X}}^{\rm{(pl)}}}}

\slugcomment{}
\shorttitle{Super-hard X-rays from V2491 Cyg}
\shortauthors{D.~Takei et al.}

\begin{document}
\title{Suzaku Detection of Super-hard X-ray Emission from the Classical Nova V2491 Cygni}
\author{
D.~Takei\altaffilmark{1},
M.~Tsujimoto\altaffilmark{2,3,4},
S.~Kitamoto\altaffilmark{1},
J.-U.~Ness\altaffilmark{4,5,6},
J.~J.~Drake\altaffilmark{7},
H.~Takahashi\altaffilmark{8}, \&
K.~Mukai\altaffilmark{9,10}
}
\email{takei@ast.rikkyo.ac.jp}

\altaffiltext{1}{Department of Physics, Rikkyo University, 3-34-1 Nishi-Ikebukuro, Toshima, Tokyo 171-8501, Japan}
\altaffiltext{2}{Japan Aerospace Exploration Agency, Institute of Space
and Astronautical Science, 3-1-1 Yoshino-dai, Sagamihara, Kanagawa 229-8510, Japan}
\altaffiltext{3}{Department of Astronomy \& Astrophysics, Pennsylvania State University, 525 Davey Laboratory, University Park, PA 16802, USA}
\altaffiltext{4}{Chandra fellow}
\altaffiltext{5}{School of Earth and Space Exploration, Arizona State University, Tempe, AZ 85287, USA}
\altaffiltext{6}{European Space Agency, XMM-Newton Observatory SOC, SRE-OAX, Apartado 78, 28691 Villanueva de la Ca\~nada, Madrid, Spain}
\altaffiltext{7}{Smithsonian Astrophysical Observatory (SAO), MS-3, 60 Garden Street, Cambridge, MA 02138, USA}
\altaffiltext{8}{Department of Physical Science, School of Science, Hiroshima
University, 1-3-1 Kagamiyama, Higashi-Hiroshima, Hiroshima 739-8526, Japan}
\altaffiltext{9}{CRESST and X-Ray Astrophysics Laboratory, NASA Goddard Space Flight Center, Greenbelt, MD 20771, USA}
\altaffiltext{10}{Department of Physics, University of Maryland, Baltimore County, 1000 Hilltop Circle, Baltimore, MD 21250, USA}

\begin{abstract}
 We report the detection of super-hard ($>$10~keV) X-ray emission extending up to 70~keV
 from the classical nova V2491 Cygni using the \textit{Suzaku} observatory. We conducted two
 $\sim$20~ks target-of-opportunity observations 9 and 29 days after the outburst on 2008
 April 11, yielding wide energy range spectra by combining the X-ray Imaging
 Spectrometer and the Hard X-ray Detector. On day 9, a spectrum was obtained at
 1.0--70~keV with the \ion{Fe}{25} K$\alpha$ line feature and a very flat continuum,
 which is explained by thermal plasma with a 3~keV temperature and power-law
 emission with a photon index of 0.1 attenuated by a heavy extinction of
 1.4$\times$10$^{23}$~cm$^{-2}$. The 15--70~keV luminosity at 10.5~kpc is
 6$\times$10$^{35}$~ergs~s$^{-1}$. The super-hard emission was not present on day
 29. This is the highest energy at which X-rays have been detected from a classical
 nova. We argue a non-thermal origin for the emission, which suggests the presence of
 accelerated charged particles in the nova explosion.
\end{abstract}

\keywords{
stars: individual (Nova Cygni 2008 number 2, V2491 Cygni)
---
stars: novae
}

\section{Introduction}
Classical nova explosions occur in accreting binaries with a white dwarf (WD) as the
primary and a late-type dwarf or giant as the secondary. Hydrogen-rich accreted material
is accumulated on the WD surface until a critical mass is reached that ignites a
thermonuclear runaway (e.g., \citealt{starrfield2008}). The released energy and mass
propagate through the circumstellar matter and are expected to form a shock structure
similar to, but in much smaller scales both in time and space than, those found in
supernova remnants (SNRs). From the analogy to SNRs, we naturally expect both thermal
and non-thermal emission in the X-ray band if we can observe classical novae with a
sufficient agility and sensitivity. Thermal emission in the hard X-ray (1--10~keV) band
from high-temperature plasma produced by adiabatic shocks has been routinely reported in
many classical novae due mainly to the \textit{Swift} satellite (e.g.,
\citealt{bode2006,ness2009}). However, no clear detection has been reported of
non-thermal X-ray emission from charged particles accelerated by shocks, as it requires
high sensitivity in the super-hard X-ray ($>$10~keV) band.

Successful detection of non-thermal X-rays will make classical novae another agent of
cosmic particle acceleration. Also, because super-hard X-ray photons can
penetrate through an extreme extinction of $\av$~$\sim$~10$^{3}$~mag, it will eventually
give a tool to unveil the currently-inaccessible phenomena shielded by a thick ejecta
material in the initial phase of classical nova explosions.

In this Letter, we report the result of the X-ray observations from the classical nova
V2491 Cyg using the \textit{Suzaku} satellite. We focus on the detection of super-hard emission
extending up to 70~keV, the highest energy X-rays ever reported from classical
novae. The results of the remaining data will be presented separately.

\section{V2491 Cyg}\label{target}
V2491 Cyg \citep{nakano2008,samus2008} was discovered on 2008 April 10.728~UT at a
Galactic coordinate of ($l$, $b$) $=$ (67.22874$^{\circ}$, 4.35315$^{\circ}$). We define
the epoch of the discovery as the origin of time in this Letter. The evolution of the
nova was extremely fast \citep{tomov2008a}, declining at a rate of $t_{2}$ $\sim$ 4.6~d
\citep{tomov2008b}, where $t_{2}$ is the time to fade by 2~mag from the optical
maximum. Using an empirical relation between the maximum magnitude and the rate of
decline among classical novae \citep{della1995}, in which intrinsically brighter ones
fade faster, a distance was estimated as $\sim$10.5~kpc \citep{helton2008}. The WD mass
is considered to be on the higher end from the rapid development of the light curves
\citep{hachisu2009}.

Intensive monitoring both by ground-based and space-based observations revealed some
distinctive characteristics of this source. In general, classical novae continue to
decline in the optical brightness, but V2491 Cyg exhibited a clear rebrightening
followed by a sudden fading around day 15. A similar rebrightening was found only in two
other cases: V1493\,Aql \citep{venturini2004} and V2362\,Cyg \citep{kimeswenger2008,lynch2008}.
\citet{hachisu2009} attempted to explain this unusual behavior by a sudden release of
magnetic energy. V2491 Cyg is also one of the few examples with X-ray detection prior to
the nova \citep{ibarra2008a,ibarra2008b,ibarra2009} with possible spectral changes among
several observations \citep{ibarra2009}. The optical rebrightening and the pre-nova
X-ray activity suggest that V2491 Cyg might host a magnetic WD.

\section{Observations and Reduction}\label{observations}
The onset of the nova triggered an X-ray monitoring campaign using the \textit{Swift}
satellite \citep{kuulkers2008,osborne2008,page2008,page2009}. No emission was found on
day 1, but X-rays emerged clearly on day 5 with extremely hard continuum emission
\citep{kuulkers2008} as well as a line-like feature near 6.7~keV attributable to
\ion{Fe}{25} K$\alpha$ in our quick look. These X-ray features are quite rare in
classical nova explosions, yet are important for understanding their high-energy
behavior. We therefore requested a $\sim$20~ks target-of-opportunity observation by
\textit{Suzaku} and obtained a well-exposed spectrum on 2008 April 19--20 (day 9.1 in
the middle of the observation). In order to follow the spectral evolution, we requested
another $\sim$20~ks time on May 9 (day 28.9). The data were processed with the pipeline
version 2.2.

\textit{Suzaku} has two instruments in operation \citep{mitsuda2007}: the X-ray Imaging
Spectrometer (XIS; \citealt{koyama2007}) sensitive at 0.2--12~keV and the Hard X-ray
Detector (HXD; \citealt{takahashi2007}; \citealt{kokubun2007}) above 10~keV. The utility
of the unique combination of detectors covering a wide energy range is illustrated by
previous \textit{Suzaku} studies of classical novae \citep{takei2008,tsujimoto2009}. 

The XIS is equipped with four X-ray CCDs at the foci of four X-ray telescope modules
\citep{serlemitsos2007}. Three of them (XIS0, 2, and 3) are front-illuminated (FI) CCDs
and the remaining one (XIS1) is back-illuminated (BI). The absolute energy scale is
accurate to $\lesssim$5~eV and the energy resolution is 160--190~eV (FWHM) at
5.9~keV. Each XIS covers a 18\arcmin$\times$18\arcmin\ field of view (FoV) with an
energy-independent half power diameter of 1\farcm8--2\farcm3. The XIS2 is not
functional, thus we used the remaining three CCDs. The XIS was operated in the normal
clocking mode with a frame time of 8~s.

The HXD is a non-imaging X-ray detector consisting of several components sensitive at
different energy ranges. We focus on the PIN detector operating at 10--70~keV with an
energy resolution of $\sim$3.0~keV. Passive fine collimators restrict the FoV to
34\arcmin$\times$34\arcmin\ (FWHM). Thanks to the surrounding anti-coincidence
scintillators, the narrow FoV, and the stable and low background environment in a low
earth orbit, the PIN achieves unprecedented sensitivity in the super-hard X-ray band.

\begin{figure}[tb]
 \epsscale{1.0}
 \plotone{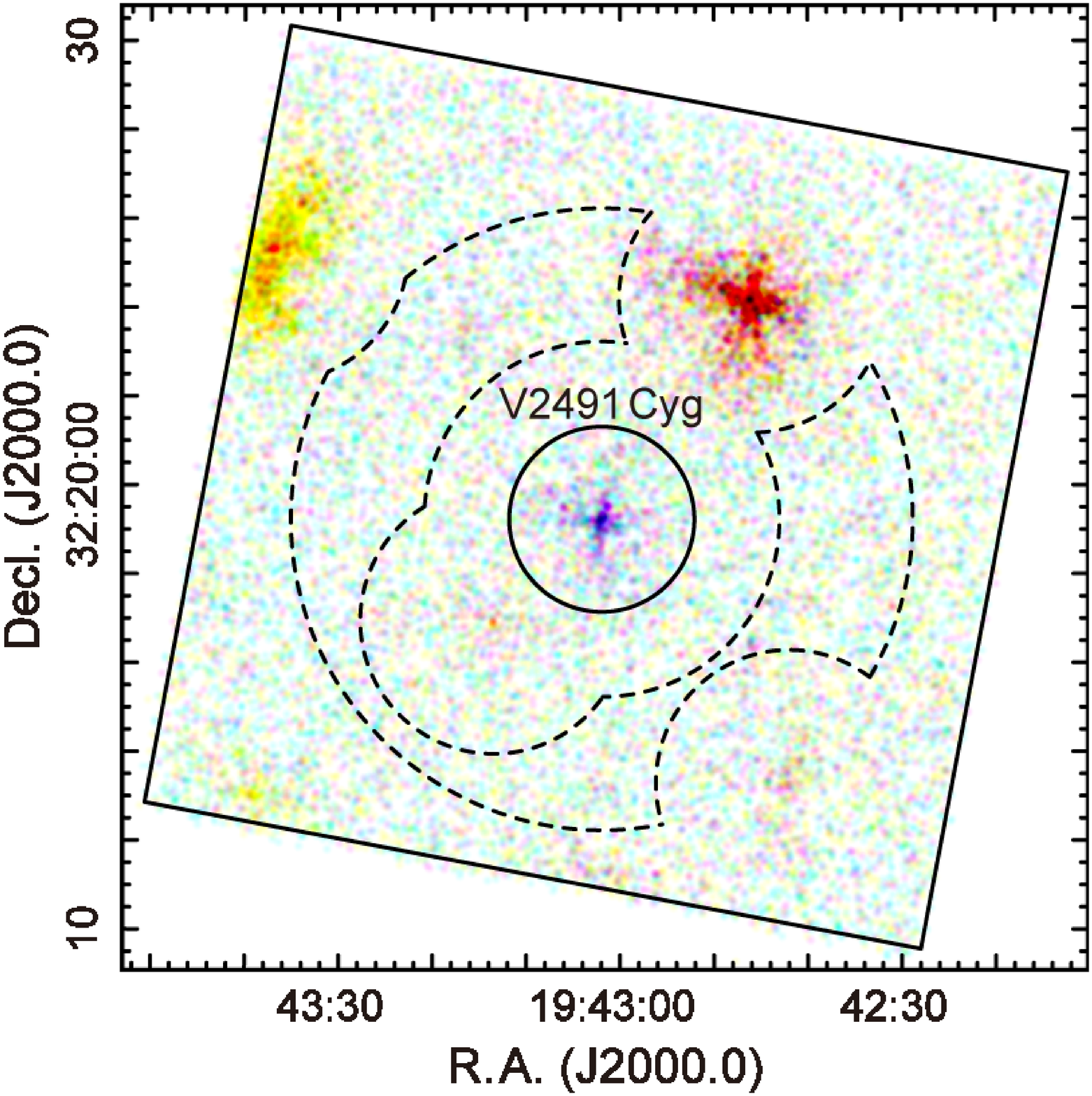}
 \caption{XIS image on day 9 coded with yellow (0.2--2.0~keV), magenta (2.0--5.0~keV),
 and cyan (7.0--12.0~keV). The source and background photons for V2491 Cyg are
 accumulated respectively from the solid and dashed regions in order to avoid
 contamination from other sources in the view.}\label{fg:image}
\end{figure}

\section{Analysis}\label{analysis}
\subsection{XIS Image and Spectrum}
\begin{figure}[tb]
 \epsscale{1.0}
 \plotone{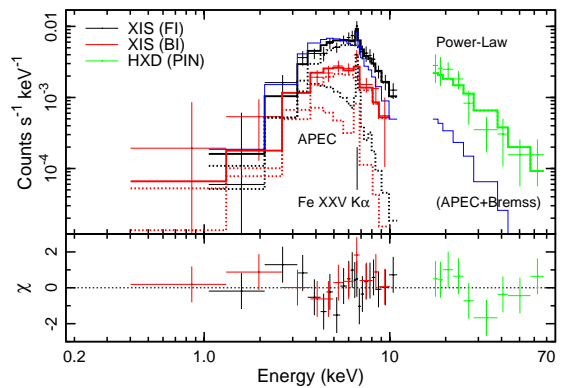}
 \caption{Background-subtracted XIS and PIN spectra on day 9 and the models. The two XIS
 FI spectra with nearly identical responses were merged, while the BI spectrum was
 treated separately. The best-fit model (APEC $+$ power-law) is shown with thick lines
 (solid for total and dashed for each component) for FI, BI and PIN in different
 colors. The alternative unsuccessful model (APEC $+$ 100 keV bremsstrahlung) is shown
 with thin blue lines for FI and PIN. The lower panel shows the residual from the
 best-fit.}\label{fg:spectrum}
\end{figure}

Figure~\ref{fg:image} shows the XIS image on day 9. V2491 Cyg is clearly detected as a
very hard source at the center with a total count rate of 4.5$\times$10$^{-2}$~s$^{-1}$
(0.2--12~keV) along with other softer sources. We constructed a background-subtracted
spectrum on day 9 (red and black symbols in Fig.~\ref{fg:spectrum}), which is
characterized by a hard flat continuum extending beyond 10~keV and a Fe K$\alpha$ line
presumably from \ion{Fe}{1} at 6.4~keV, \ion{Fe}{25} at 6.7~keV, or \ion{Fe}{26} at
7.0~keV between 6 and 7~keV. No emission in the soft band is visible. On day 29, the
spectrum became much softer, and the hard continuum and the Fe K$\alpha$ line emission
disappeared. The detailed results of the second spectrum are presented in Takei et
al. in prep.

\subsection{PIN Spectrum}
We see a high prospect of PIN detection on day 9, in which the XIS spectrum shows very
flat continuum emission extending beyond 10~keV. We examine the signal against the
background, which is a combination of the instrumental non-X-ray background (NXB) and
the X-ray background. The X-ray background consists of the Cosmic X-ray background (CXB)
emission and the Galactic ridge X-ray emission (GRXE). The majority of the background is
the NXB; the CXB is $\sim$2 orders smaller than the NXB, and the GRXE is totally
negligible with $\gtrsim$1 order smaller than the CXB toward V2491 Cyg
\citep{revnivtsev2006}.

\begin{figure}[tb]
 \epsscale{1.0}
 \plotone{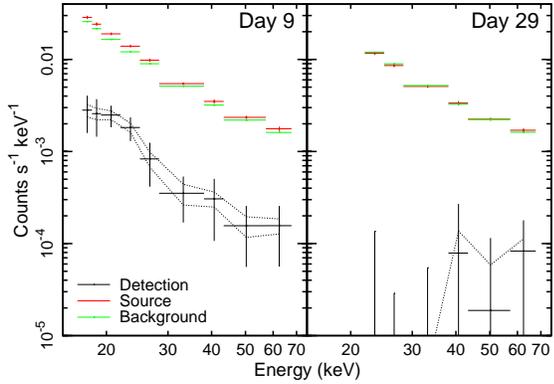}
 \caption{Comparison of the observed signal (red) against the model background (green)
 of the PIN data on days 9 (left) and 29 (right). The background includes both the NXB
 and the CXB. The background-subtracted spectrum is shown in black. The dotted band
 indicates the uncertainty stemming from the inaccuracy of the reproducibility in the
 NXB model. On day 29, the data are not used below 20~keV due to high detector
 temperatures.}\label{fg:spectrum_hxd}
\end{figure}

In Figure~\ref{fg:spectrum_hxd}, we compare the observed signal against the model
background. The NXB model was simulated by taking account of the cut-off-rigidity
history and the elapsed time from South Atlantic anomaly passages during the
observation. The 1~$\sigma$ reproducibility of the NXB is $\lesssim$1.8~\% in flux
\citep{fukazawa2009}. The CXB model was constructed by convolving the detector responses
with the CXB model derived by \textit{HEAO-1} observations \citep{boldt1987}. A
significant signal up to 70~keV was detected at a $\gtrsim$3~$\sigma$ level on day
9. The detection is also significant even when subtracting the maximum allowable NXB
model. The detection is further confirmed by subtracting another NXB model constructed
differently by accumulating the PIN data of our own observation during the earth
occultation. In contrast, no significant PIN emission was detected on day 29.

We attribute the PIN emission on day 9 to V2491 Cyg for the following reasons: (1) the
flux of the PIN emission is consistent with the extrapolated flux of the flat hard
continuum in the hard end of the XIS spectrum (Fig.~\ref{fg:spectrum}), (2) no other
X-ray source with a comparable hardness was found in the XIS image
(Fig.~\ref{fg:image}), (3) no super-hard X-ray source is known within the PIN FoV by the
\textit{INTEGRAL} \citep{ebisawa2003} and the quick-look data by the All-Sky Monitor
in the \textit{Rossi X-ray Timing Explorer} (\textit{RXTE}), and (4) the absence of the PIN
emission on day 29 is consistent with the fact that the extrapolated XIS emission is too
weak for any PIN detection.

\subsection{Spectral Fitting}
We now fit the resultant background-subtracted XIS and PIN spectra on day 9
simultaneously. We start with the Fe K$\alpha$ feature between 6 and 7~keV. We fitted
the feature locally with a Gaussian line and obtained the best-fit energy and equivalent
width of 6.6$\pm$0.1~keV and $\sim$240~eV, respectively. The detection of the line was
found significant at a $>$3~$\sigma$ level both by the F-test and a Monte Carlo
simulation \citep{protassov2002}. We attribute the feature to the \ion{Fe}{25} K$\alpha$
emission line at 6.7~keV. We further examined the presence of possible accompanying
lines at 6.4~keV by \ion{Fe}{1} K$\alpha$ and at 7.0~keV by \ion{Fe}{26} K$\alpha$, but
these lines were found insignificant.

The presence of the \ion{Fe}{25} K$\alpha$ line and the absence of the \ion{Fe}{26}
K$\alpha$ line indicate a thermal origin of the emission by plasma with a temperature of
$\lesssim$10~keV. The plasma at this temperature is incapable of producing the flat
super-hard continuum, so an additional harder spectral component is required. We fitted
the entire spectrum with two different combinations of models. The first one is
optically-thin thermal plasma (APEC; \citealt{smith2001}) plus a power-law model, which
is respectively for the thermal features in the hard band and the flat continuum across
both bands. The second combination employs a bremsstrahlung model in place of the
power-law model. These models are attenuated by an interstellar extinction model (TBabs;
\citealt{wilms2000}). The Fe abundance relative to H in the plasma model was not
constrained well by the data, and we therefore fixed it at solar \citep{anders1989}.

We obtained a statistically acceptable fit for the first combination. The best-fit
parameter values are compiled in Table~\ref{tb:parameter01} for the amount of extinction
($\nh$), the thermal plasma temperature and flux ($\ktFT$ and $\fxFT$), and the
power-law photon index and flux ($\pwFP$ and $\fxFP$). The extinction-corrected
luminosity at 10.5~kpc was also derived for each component ($\lxFT$ and $\lxFP$). For the second
combination of models, the electron temperature of the bremsstrahlung emission is
constrained to be $>$100~keV, which is much higher than the post shock temperature in a
strong shock by the ejecta propagating at $\sim$4000~km~s$^{-1}$ (e.g.,
\citealt{tomov2008a,tomov2008b}). We therefore conclude that the spectrum on day 9 is
better explained by a combination of 3~keV thermal plasma and power-law emission with a
photon index of 0.1 attenuated by a heavy extinction. The systematic uncertainty in the
NXB reproducibility for the PIN brings little change in the result by about $\pm$0.3~keV
in $\ktFT$, $\pm$0.08 in $\pwFP$, and $\pm$15\% both in $\fxFT$ and $\fxFP$.

\begin{table}[tb]
 \vspace{-0.2cm}
 \begin{center}
  \caption{Best-fit parameters on day 9.}\label{tb:parameter01}
  \begin{tabular}{llllll}
   \tableline
   Comp.      & Par.                     & Unit                      & Value\tablenotemark{a} \\
   \tableline
   Absorption & \nh                      & (cm$^{-2}$)               & 1.4$_{-0.4}^{+0.9}\times$10$^{23}$ \\
   Power-law  & \pwFP                    &                           & 0.1$_{-0.2}^{+0.2}$ \\
              & \fxFP\tablenotemark{b}   & (ergs~s$^{-1}$~cm$^{-2}$) & 4.8$_{-1.3}^{+0.7}\times$10$^{-11}$ \\
              & \lxFP\tablenotemark{b}   & (ergs~s$^{-1}$)           & 6.4$_{-1.7}^{+0.9}\times$10$^{35}$  \\
   Thermal    & \ktFT                    & (keV)                     & 2.9$_{-2.6}^{+4.3}$ \\
              & \fxFT\tablenotemark{b}   & (ergs~s$^{-1}$~cm$^{-2}$) & 1.4$_{-1.1}^{+12}\times$10$^{-13}$ \\
              & \lxFT\tablenotemark{b}   & (ergs~s$^{-1}$)           & 1.9$_{-1.4}^{+16}\times$10$^{33}$ \\
   \tableline
   \multicolumn{3}{l}{$\chi^{2}/\rm{d.o.f.}$} & \multicolumn{1}{l}{22.8/35} \\
   \tableline
  \end{tabular}
  \vspace{-0.4cm}
  \tablenotetext{a}{The statistical uncertainties indicate the 90\% confidence ranges.}
  \tablenotetext{b}{The values are in 1.0--12.0 and 15--70~keV respectively for the
  thermal and power-law emission. The luminosities are at 10.5~kpc.}
 \end{center}
\end{table}

\section{Discussion}\label{discussion}
We have found the super-hard X-ray emission extending up to 70~keV from V2491 Cyg on day
9. The spectrum cannot be explained by a simple bremsstrahlung model with a reasonable
temperature. The best-fit power-law model is very flat with a photon index of 0.1$\pm$0.2.
The emission was absent on day 29 in an observation with the same exposure. The
15--70~keV flux decreased from 4.8$\times$10$^{-11}$~ergs~s$^{-1}$~cm$^{-2}$ on day 9 to
$\lesssim$1.5$\times$10$^{-11}$~ergs~s$^{-1}$~cm$^{-2}$ on day 29 (a 90\% confidence
upper limit assuming the same spectral shape). The rapid decay rules out the possibility
of Comptonization of nuclear $\gamma$-ray lines from radioactive species such as
$^{22}$Na \citep{livio1992} or rekindled accretion \citep{hernanz02}.

Super-hard X-ray emission was studied in two other classical novae to date; V382\,Vel by
\textit{BeppoSAX} \citep{orio2001}, and RS\,Oph by \textit{RXTE} \citep{sokolosky2006}
and \textit{Swift} \citep{bode2006}. Similarly to V2491 Cyg, these observations were
conducted a short time after the outburst (days 3 and 15 for RS\,Oph and V382\,Vel,
respectively). However, neither of the two other cases shows clear evidence of
non-thermal signature of the emission. In RS\,Oph, \textit{RXTE} detected photons only
below $\sim$20~keV, which is attributable to thermal plasma with a reasonable
temperature of $\lesssim$10~keV \citep{sokolosky2006}. \textit{Swift} reported weak
25--50~keV detection but no spectrum was obtained in the super-hard X-ray band
\citep{bode2006}. In V382\,Vel, \textit{BeppoSAX} only constrained a 2~$\sigma$ upper
limit in the 15--60~keV band \citep{orio2001}. On the contrary, in V2491 Cyg, the
spectrum is clearly seen up to 70~keV, which is not explained by a simple bremsstrahlung
model with a reasonable temperature but by an extremely flat power-law model.

The power-law emission from V2491 Cyg suggests the presence of an accelerated population
of electrons with non-thermal energy distribution. Non-thermal particles in classical
nova explosions are suggested in some radio observations and theoretical studies of
RS\,Oph. \citet{rupen2008} found evidence of synchrotron emission in a high-resolution
radio interferometer observation. \citet{tatischeff2007} argued for the presence of
diffusive shock acceleration based on the fact that the blast wave decelerated at a much
faster rate than that predicted by the standard adiabatic shock model. Our result is the
first to claim a non-thermal signature from classical nova explosions in the X-ray band.

The extremely flat photon index of 0.1$\pm$0.2 in the power-law radiation poses a
challenge to understand the radiation mechanism as well as the acceleration mechanism of
electrons to produce the radiation. If the dominant radiation is inverse Compton or
synchrotron emission and the electrons lose energy radiatively, the electron population
has a number index from $-$0.8 to $-$1.8 in the power-law energy distribution at
injection. If bremsstrahlung emission is the dominant radiation and the
non-relativistic electrons lose energy via ionization losses with ambient matter, the
injected electrons have a number index from $-$0.4 to 1.1. These values are too hard for
standard diffusive shock acceleration, in which a number index of 2.0 is expected
\citep{blandford1980}. Additional explanations, such as multiple scattering in the
inverse Compton process or the acceleration by magnetic reconnections, would be
necessary.

Alternative interpretations to the non-thermal model for the super-hard emission include
a multiple temperature hot ($\gtrsim$10~keV) plasma with an extreme extinction
($\gtrsim$10$^{25}$~cm$^{-2}$) and Comptonized blackbody emission
\citep{nishimura1986}. The former is unlikely due to the lack of reprocessed emission
(e.g., \ion{Fe}{1} K$\alpha$) and the fact that such a large extinction cannot be
realized by the typical amount of ejecta ($\sim$10$^{-5}$~$M_{\odot}$; e.g.,
\citealt{pietsch2007}). The latter requires blackbody emission of $\sim$5~keV, which is
too hot for the WD surface emission.

\vspace{-0.5cm}
\acknowledgments

The authors appreciate the reviewer, Klaus Beuermann, for useful suggestions. We also
thank the \textit{Suzaku} telescope managers for the director's discretionary time and
I. Hachisu, M. Kato, K. Kinugasa, H. Murakami, and M. Morii for comments. D.\,T. is
financially supported by the Japan Society for the Promotion of Science. Support for
this work was provided by the NASA through \textit{Chandra} Postdoctoral Fellowship
Awards (PF6-70044; M.\,T. and PF5-60039; J.-U.\,N.) operated by the SAO for and on
behalf of the NASA under contract NAS8-03060. J.\,J.\,D. was supported by the NASA
contract NAS8-39073.

\end{document}